\input harvmac.tex
\noblackbox
\input epsf.sty

\font\cmss=cmss10
\font\cmsss=cmss10 at 7pt

\def\inbar{\vrule height1.5ex width.4pt depth0pt}

\def\IN{\relax{\rm I\kern-.18em N}}
\def\IB{\relax\hbox{$\inbar\kern-.3em{\rm B}$}}
\def\IC{\relax\hbox{$\inbar\kern-.3em{\rm C}$}}
\def\IQ{\relax\hbox{$\inbar\kern-.3em{\rm Q}$}}
\def\ID{\relax\hbox{$\inbar\kern-.3em{\rm D}$}}
\def\IE{\relax\hbox{$\inbar\kern-.3em{\rm E}$}}
\def\IF{\relax\hbox{$\inbar\kern-.3em{\rm F}$}}
\def\IG{\relax\hbox{$\inbar\kern-.3em{\rm G}$}}
\def\IGa{\relax\hbox{${\rm I}\kern-.18em\Gamma$}}
\def\IH{\relax{\rm I\kern-.18em H}}
\def\IK{\relax{\rm I\kern-.18em K}}
\def\IL{\relax{\rm I\kern-.18em L}}
\def\IP{\relax{\rm I\kern-.18em P}}
\def\IR{\relax{\rm I\kern-.18em R}}
\def\Z{\relax\ifmmode\mathchoice{\hbox{\cmss Z\kern-.4em Z}}{\hbox{\cmss Z\kern-.4em Z}} {\lower.9pt\hbox{\cmsss Z\kern-.4em Z}}{\lower1.2pt\hbox{\cmsss Z\kern-.4em Z}}\else{\cmss Z\kern-.4em Z}\fi}
\def\IZ{Z\!\!\!Z}
\def\II{\relax{\rm I\kern-.18em I}}
\def\one{\relax{\rm 1\kern-.25em I}}

\def\CLL{\relax{\CL\kern-.74em \CL}}

\def\CA{{\cal A}}

\def\CF{{\cal F}}

\def\CN{{\cal N}}
\def\CO{{\cal O}}

\def\CS{{\cal S}}

\def\CV{{\cal V}}
\def\CW{{\cal W}}

\lref\Verlinde{
  M.~Buican, D.~Malyshev and H.~Verlinde,
  ``On the Geometry of Metastable Supersymmetry Breaking,''
  arXiv:0710.5519 [hep-th].
}
\lref\HM{
  J.~A.~Harvey and G.~W.~Moore,
  ``On the algebras of BPS states,''
  Commun.\ Math.\ Phys.\  {\bf 197}, 489 (1998)
  [arXiv:hep-th/9609017].
}
\lref\ABF{
  M.~Aganagic, C.~Beem and B.~Freivogel,
  ``Geometric Metastability, Quivers and Holography,''
  arXiv:0708.0596 [hep-th].
}
\lref\ABSV{
  M.~Aganagic, C.~Beem, J.~Seo and C.~Vafa,
  ``Geometrically induced metastability and holography,''
  arXiv:hep-th/0610249.
}
\lref\VUD{
  F.~Cachazo, B.~Fiol, K.~A.~Intriligator, S.~Katz and C.~Vafa,
  ``A geometric unification of dualities,''
  Nucl.\ Phys.\  B {\bf 628}, 3 (2002)
  [arXiv:hep-th/0110028].
}
\lref\ABK{
  M.~Aganagic, C.~Beem and S.~Kachru,
  ``Geometric Transitions and Dynamical SUSY Breaking,''
  arXiv:0709.4277 [hep-th].
}
\lref\CIV{
  F.~Cachazo, K.~A.~Intriligator and C.~Vafa,
  ``A large N duality via a geometric transition,''
  Nucl.\ Phys.\  B {\bf 603}, 3 (2001)
  [arXiv:hep-th/0103067].
}
\lref\CKV{
  F.~Cachazo, S.~Katz and C.~Vafa,
  ``Geometric transitions and N = 1 quiver theories,''
  arXiv:hep-th/0108120.
}
\lref\PolchinskiSM{
  J.~Polchinski and A.~Strominger,
  ``New Vacua for Type II String Theory,''
  Phys.\ Lett.\  B {\bf 388}, 736 (1996)
  [arXiv:hep-th/9510227].
}
\lref\LM{
  A.~Lawrence and J.~McGreevy,
  ``Local string models of soft supersymmetry breaking,''
  JHEP {\bf 0406}, 007 (2004)
  [arXiv:hep-th/0401034].
}
\lref\Hitchin{
  M.~Grana, J.~Louis and D.~Waldram,
  ``Hitchin functionals in N = 2 supergravity,''
  JHEP {\bf 0601}, 008 (2006)
  [arXiv:hep-th/0505264].
}
\lref\Mich{
  J.~Michelson, ``Compactifications of type IIB strings to four
  dimensions with non-trivial classical potential,'' Nucl.\ Phys.\ B
  {\bf 495}, 127 (1997) [arXiv:hep-th/9610151].  
}
\lref\APT{
  I.~Antoniadis, H.~Partouche and T.~R.~Taylor,
  ``Spontaneous Breaking of N=2 Global Supersymmetry,''
  Phys.\ Lett.\  B {\bf 372}, 83 (1996)
  [arXiv:hep-th/9512006].
}
\lref\IZ{
  E.~A.~Ivanov and B.~M.~Zupnik,
  ``Modified N = 2 supersymmetry and Fayet-Iliopoulos terms,''
  Phys.\ Atom.\ Nucl.\  {\bf 62}, 1043 (1999)
  [Yad.\ Fiz.\  {\bf 62}, 1110 (1999)]
  [arXiv:hep-th/9710236].
}
\lref\TV{
  T.~R.~Taylor and C.~Vafa,
  ``RR flux on Calabi-Yau and partial supersymmetry breaking,''
  Phys.\ Lett.\  B {\bf 474}, 130 (2000)
  [arXiv:hep-th/9912152].
}
\lref\BD{
  M.~Berkooz, M.~R.~Douglas and R.~G.~Leigh,
  ``Branes intersecting at angles,''
  Nucl.\ Phys.\  B {\bf 480}, 265 (1996)
  [arXiv:hep-th/9606139].
}
\lref\AhK{
  O.~Aharony and S.~Kachru,
  ``Stringy Instantons and Cascading Quivers,''
  JHEP {\bf 0709}, 060 (2007)
  [arXiv:0707.3126 [hep-th]].
}
\lref\AKS{
  O.~Aharony, S.~Kachru and E.~Silverstein,
  ``Simple Stringy Dynamical SUSY Breaking,''
  arXiv:0708.0493 [hep-th].
}
\lref\SW{
  N.~Seiberg and E.~Witten,
  ``String theory and noncommutative geometry,''
  JHEP {\bf 9909}, 032 (1999)
  [arXiv:hep-th/9908142].
}
\lref\PS{
  J.~Polchinski and A.~Strominger,
  ``New Vacua for Type II String Theory,''
  Phys.\ Lett.\  B {\bf 388}, 736 (1996)
  [arXiv:hep-th/9510227].
}
\lref\V{
  C.~Vafa,
  ``Superstrings and topological strings at large N,''
  J.\ Math.\ Phys.\  {\bf 42}, 2798 (2001)
  [arXiv:hep-th/0008142].
}
\lref\DDD{
  D.~E.~Diaconescu, M.~R.~Douglas and J.~Gomis,
  ``Fractional branes and wrapped branes,''
  JHEP {\bf 9802}, 013 (1998)
  [arXiv:hep-th/9712230].
}
\lref\DM{
  M.~R.~Douglas and G.~W.~Moore,
  ``D-branes, Quivers, and ALE Instantons,''
  arXiv:hep-th/9603167.
}
\lref\CV{
  F.~Cachazo and C.~Vafa,
  ``N = 1 and N = 2 geometry from fluxes,''
  arXiv:hep-th/0206017.
}
\lref\D{
  M.~R.~Douglas,
  ``Enhanced gauge symmetry in M(atrix) theory,''
  JHEP {\bf 9707}, 004 (1997)
  [arXiv:hep-th/9612126].
}
\lref\AganagicNN{
  M.~Aganagic, C.~Popescu and J.~H.~Schwarz,
  ``Gauge-invariant and gauge-fixed D-brane actions,''
  Nucl.\ Phys.\  B {\bf 495}, 99 (1997)
  [arXiv:hep-th/9612080].
}
\lref\Yu{
  Y.~Nakayama, M.~Yamazaki and T.~T.~Yanagida,
  ``Moduli Stabilization in Stringy ISS Models,''
  arXiv:0710.0001 [hep-th].
}
\lref\warped{
  M.~R.~Douglas, J.~Shelton and G.~Torroba,
  ``Warping and supersymmetry breaking,''
  arXiv:0704.4001 [hep-th].
}

\lref\CamaraCZ{
  P.~G.~Camara and M.~Grana,
  ``No-scale supersymmetry breaking vacua and soft terms with torsion,''
  arXiv:0710.4577 [hep-th].
}
\lref\granatwo{
  A.~Butti, M.~Grana, R.~Minasian, M.~Petrini and A.~Zaffaroni,
  ``The baryonic branch of Klebanov-Strassler solution: A supersymmetric
  family of SU(3) structure backgrounds,''
  JHEP {\bf 0503}, 069 (2005)
  [arXiv:hep-th/0412187].
}

\lref\dymarsky{
  A.~Dymarsky, I.~R.~Klebanov and N.~Seiberg,
  ``On the moduli space of the cascading SU(M+p) x SU(p) gauge theory,''
  JHEP {\bf 0601}, 155 (2006)
  [arXiv:hep-th/0511254].
}
\lref\emanuel{
  D.~E.~Diaconescu, A.~Garcia-Raboso and K.~Sinha,
  ``A D-brane landscape on Calabi-Yau manifolds,''
  JHEP {\bf 0606}, 058 (2006)
  [arXiv:hep-th/0602138].
}


\Title{\vbox{\baselineskip12pt\hbox{}
\hbox{{UCB-PTH-07/22}}
\hbox{}}}
{\vbox{ {\centerline{Geometric Transitions and D-term SUSY Breaking} }}}

\centerline{Mina Aganagic and  Christopher Beem}
\bigskip
\centerline{{\it Department of Physics, University of California, Berkeley, CA 94720}}

\bigskip
\bigskip
\bigskip
\bigskip

\noindent
We propose a new way of using geometric transitions to study
metastable vacua in string theory and certain confining gauge theories.
The gauge theories in question are 
${\cal N}=2$ supersymmetric theories deformed to ${\cal N}=1$ by superpotential terms. We
first geometrically engineer supersymmetry-breaking vacua by wrapping
D5 branes on rigid 2-cycles in noncompact Calabi-Yau geometries, such
that the central charges of the branes are misaligned. In a limit of slightly misaligned charges,
this has a gauge theory description, where supersymmetry is broken by Fayet-Iliopoulos D-terms. 
Geometric
transitions relate these configurations to dual Calabi-Yaus
with fluxes, where $H_{RR}$, $H_{NS}$ and $dJ$ are all nonvanishing.  
We argue that the dual geometry can be effectively used
to study the resulting non-supersymmetric, confining vacua.

\Date{November 2007}

\newsec{Introduction}

In the interest of finding controllable, realistic string vacua, it is
important to find simple and tractable mechanisms of breaking
supersymmetry in string theory. A powerful method which has been put
forward in \ABSV\ consists of geometrically engineering metastable
vacua with D-branes wrapping cycles in a Calabi-Yau manifold, and
using geometric transitions and topological string techniques to
analyze them. In \ABSV, metastable vacua were engineered by wrapping
D5 branes and anti-D5 branes on rigid 2-cycles in a Calabi-Yau. The
non-supersymmetric vacuum obtained in this fashion was argued to have
a simple closed-string dual description, in which branes and
antibranes are replaced by fluxes.  In \ABK, geometric transitions
were used to study the physics of D-brane theories that break
supersymmetry dynamically.  Namely, the authors showed that the
instanton generated superpotential that triggers supersymmetry
breaking can be computed by classical means in a dual geometry, where some of
the branes are replaced by fluxes (For an alternative approach see \AhK).

In this paper, we propose another way to use geometric transitions
to study supersymmetry breaking. As in \ABSV, we consider D5 branes
wrapping rigid cycles in a noncompact Calabi-Yau $X$. 
If $b_2(X)>1$, supersymmetry can be broken by choosing the complexified K\"ahler moduli so as to misalign the central charges of the branes,
$$
Z_i  = \int_{S^2_i} J+ i B_{NS}.
$$
Since the 2-cycles wrapped by the D5 branes are rigid, any deformation of
the branes costs energy, and the system is guaranteed to be
metastable.\foot{Supersymmetry breaking by missaligning the central charges of D5 branes wrapped on rigid curves was also studied in \emanuel , in the context of compact Calabi-Yau manifolds.} 
In the extreme case of anti-aligned central charges, we
recover the brane/antibrane configurations of \refs{\ABSV, \ABF}. For
slightly misaligned central charges, the system has a gauge theory
description in terms of an ${\cal N}=2$ quiver theory deformed
to ${\cal N}=1$ by superpotential terms \refs{\CIV,\CKV}, and with
Fayet-Iliopoulos D-terms turned on \refs{\DM,\DDD,\D}.\foot{Some geometric aspects of supersymmetry breaking by F-terms in this context were recently discussed in \Verlinde.} The latter
trigger spontaneous supersymmetry breaking in the gauge theory.

We argue that the dynamics of this system is effectively
captured by a dual Calabi-Yau with all branes replaced by fluxes. Turning
on generic K\"ahler moduli on the open-string side has a simple
interpretation in the dual low-energy effective theory as turning on a
more generic set of FI parameters than hitherto considered in this context, but which
are allowed by the ${\cal N}=2$ supersymmetry of the
background. On-shell, this breaks some or all of the ${\cal N}=2$
supersymmetry. Geometrically, this corresponds to not only turning on
$H_{NS}$ and $H_{RR}$ fluxes on the Calabi-Yau, but also allowing for 
$dJ\neq 0$ \refs{\LM,\Hitchin}.\foot{For another example of supersymmetry breaking by turning on $H_{NS}$, $H_{RR}$ and $dJ$ fluxes, see \CamaraCZ .} 
Moreover, we show that the Calabi-Yau geometries
with these fluxes turned on have non-supersymmetric, metastable vacua,
as expected by construction in the open-string theory.

The paper is organized as follows: In section two, we review the
physics of D5 branes on a single conifold and the dual geometry after
the transition, paying close attention to the effect of
Fayet-Iliopoulos terms.  In section three, we consider the case of an
$A_2$ geometry where misaligned central charges lead to supersymmetry
breaking.  We provide evidence that the dual geometry correctly
captures the physics of the non-supersymmetric brane system. We show
that the results are consistent with expectations from the gauge theory, 
to the extent that these are available. We also comment
on the relation of this work to \ABK, and point out some possible future directions. 
In an appendix, we lay out the
general case for larger quiver theories. We show that in the limit of large separations between nodes, the theory has metastable, non-supersymmetric vacua in all cases where they are expected.

\newsec{The Conifold}

In this section, we consider $N$ D5 branes on the resolved conifold.  We will first review the open-string theory on the branes, and then discuss the dual closed-string description.  Our discussion will be more general than the canonical treatment in that we will consider the case where the D5 branes possess an arbitrary central charge.

\subsec{The D-brane theory}

To begin with, let us recall the well-known physics of $N$ D5 branes wrapping the $S^2$ tip of the resolved conifold.  This geometry can be represented as a hypersurface in $\;\IC^4[u,v,z,t]$,
$$
uv=z(z-mt).
$$
The geometry has a singularity at the origin of ${\;\IC^4}$ which
can be repaired by blowing up a rigid $\IP^1$.  This gives the ${\IP}^1$ a 
complexified K\"ahler class
\eqn\cntrl{
Z=\int_{S^2}(J+i B_{NS}) = j+i\, b_{NS}.  }
The theory on the D5 branes at vanishing $j$ reduces in the field
theory limit to a $d=4$, ${\cal N}=1$, $U(N)$ gauge theory with an
adjoint valued chiral superfield of mass $m$. The bare gauge coupling
is given by
\eqn\first{
{4\pi\over g_{{YM}}^2}={b_{NS}\over g_s},
}
for positive $b_{NS}$.
In string theory, the tension of the branes generates an energy density
related to the four dimensional gauge coupling 
by
\eqn\tvj{V_{*} =~{2N\over g_{YM}^2}=
~N{b_{NS}\over 2\pi g_s}.
}

Turning on a small, nonzero $j$ can be viewed as a 
deformation of this theory by a Fayet-Iliopoulos parameter 
for the $U(1)$ center of the gauge group \refs{\DM,\DDD}. This deforms 
the Lagrangian by
\eqn\DD{
\Delta\CL={\sqrt 2} \xi\, {\rm Tr} D
}
where 
$D$ is the auxiliary field in the ${\cal N}=1$ vector 
multiplet and
\eqn\FI{
\xi = {j\over 4\pi g_s},
}
where the factor of $g_s$ comes from the disk amplitude.
This deformation \DD\ breaks the ${\cal N}=1$ supersymmetry which was 
linearly realized at $j=0$.
In particular, turning on $j$, increases the energy of the vacuum:
integrating out $D$ from the theory by completing the square in the auxiliary field Lagrangian,
$$
\CL_D = {1 \over 2g_{YM}^2} {\rm Tr} D^2 + \sqrt{2} \xi {\rm Tr} D,
$$
raises the vacuum energy to
\eqn\new{
V_{*} = ~N {b_{NS}\over 2 \pi g_s}\Bigl(1 +{1\over 2} {j^2\over b_{NS}^2}\Bigr).
}

Supersymmetry is {\it not} broken however. At nonzero $j$, a different
${\cal N}=1$ supersymmetry is preserved\foot{This is true even in the
field theory limit, despite the presence of the constant FI
term. Namely, a second, nonlinearly realized supersymmetry is
present in the gauge theory as long as there is only a constant energy
density \HM . We thank A. Strominger for explaining this to us.}  --
one that was realized nonlinearly at vanishing $j$
\refs{\HM,\AganagicNN}. Which subgroup of the background ${\cal N}=2$
supersymmetry is preserved by the branes is determined by $Z$ in
\cntrl, the BPS central charge in the extended supersymmetry
algebra.\foot{Strictly speaking, the central charge of $N$ branes is
$NZ$. In this paper, we will always take the number of branes $N$ to be
positive, so that we interpolate between the branes and antibranes
by varying $Z$.}  For any $Z$, the open-string theory on the branes
has an alternative description which is manifestly ${\cal N}=1$
supersymmetric, with $vanishing$ FI term and with a bare gauge
coupling related to the magnitude of the central charge \CKV,
\eqn\second{
{1\over {\tilde g}_{{YM}}^2}=
{\sqrt{b_{NS}^2+j^2}\over 4 \pi g_s}.
}
Geometrically, this is just the quantum volume of the resolving $\IP^1$. As such, the central charge also determines the exact tension of a single D5 brane at nonzero $j$, so
\eqn\ven{
V_{*} = N
{\sqrt{b_{NS}^2+j^2}\over 2 \pi g_s}.
}
For small K\"ahler parameter,
$$
j \ll b_{NS},
$$ 
this agrees with the vacuum energy in the field theory limit \new.

For any $j$, the theory is massive; it is expected to exhibit confinement and gaugino condensation at low energies, leaving an effective $U(1)$ gauge theory 
in terms of the center of the original $U(N)$ gauge group. We'll show next 
that the strongly coupled theory has a simple description for $any$ value of $Z$  in terms of a large $N$ dual geometry with fluxes.

\subsec{The geometric transition at general $Z$}

We'll now discuss the large $N$ dual geometry for general values of
the central charge $Z$. Special cases (either vanishing $j$ or
vanishing $b_{NS}$) have been considered in the literature, but the
present, expanded discussion is, to our knowledge, new.\foot{See related discussion
in \CKV .} We'll see that the dual geometry exactly reproduces the
expected D5 brane physics. From the perspective of the low-energy
effective action, the consideration of general central charge
corresponds to turning on a more general set of ${\cal N}=2$ FI terms
than previously considered in this context.  Geometrically, this will
lead us to consider generalized Calabi-Yau manifolds, for which $dJ$ is
nonvanishing in addition to having $H_{NS}$ and $H_{RR}$ fluxes turned
on.  This will provide a local description of the physics for each set
of branes in the more general supersymmetry-breaking cases of sections
three and four.

To begin, let us recall the large $N$ dual description of the D5 brane
theory at vanishing $j$. This is given in terms of closed-string theory on the deformed conifold geometry,
\eqn\coni{
uv=z(z-mt)+s.
}
This is related to the open-string geometry by a geometric transition which shrinks the ${\IP}^1$ and replaces it with an $S^3$
of nonzero size,
$$
S = \int_{A} \Omega,
$$
where $A$ is the 3-cycle corresponding to the new $S^3$, and the
period of the holomorphic three-form over $A$ is related to the
parameters of the geometry by $S = s/m$. The D5 branes have
disappeared and have been replaced by $N$ units of Ramond-Ramond flux
through the $S^3$,
\eqn\RR{ \int_{A} H^{RR} = N.
}
There are Ramond-Ramond and Neveu-Schwarz fluxes through the dual, noncompact $B$-cycle as well,
\eqn\oth{
\alpha = \int_{B}^{\Lambda_0}( H_{RR} + i{H_{NS}/ g_s}) = b_{RR}+ {i} {b_{NS}/ g_s},
}
which corresponds to the complexified gauge coupling of the open-string theory,
$$
\alpha = {\theta\over2\pi} + {4\pi i \over g_{YM}^2}.
$$
The $B$-cycle is cut off at the scale, $\Lambda_0$, at which $\alpha$ 
is measured.\foot{For simplicity, the IIB axion is set to zero in this paper.}  The dependence of $\alpha$ on the IR cutoff in the geometry corresponds to its renormalization group running in the open-string theory.  
  
If it were not for the fluxes, the theory would have ${\cal N}=2$
supersymmetry, with $S$ being the lowest component of an ${\cal N}=2$
$U(1)$ vector multiplet. That theory is completely described by specifying the prepotential, ${\cal F}_0(S),$ which can be determined by a classical geometry computation,
$$
\int_{B} \Omega  = {\del\over \del {S}} {\cal F}_0.
$$ 
The presence of nonzero fluxes introduces electric and magnetic
Fayet-Iliopoulos terms in the low-energy theory for the $U(1)$ vector 
multiplet and its magnetic dual \refs{\TV, \V, \APT,\IZ}. 
The effect of the fluxes \RR,\oth\ can also be described in the language of $\CN=1$ superspace as turning on a 
superpotential for the ${\cal N}=1$ chiral superfield with $S$ as
its scalar component,
$$
\CW(S)=\int_X  \Omega\wedge(H^{RR} + {i H^{NS}/ g_s}).
$$
For the background in question, this takes the form
\eqn\suppp{
\CW(S)= \alpha S - N {\del\over \del S} {\cal F}_0.
}
In terms of the parameters of the D-brane theory, $S$ is 
identified with the vev of the gaugino condensate. One way to see this is by
comparing the superpotentials on the two sides of the duality. The  
$\alpha S$ superpotential on the closed string side 
corresponds to the classical superpotential term 
${\alpha\over 4} Tr W_{\alpha} W^{\alpha}$ on the gauge theory side.

What does the FI term deformation of the D-brane theory correspond 
to in the closed-string theory?  To begin with, let us address this question 
from the perspective of the low-energy effective action.  We know that 
the $U(1)$ gauge field after the transition coincides \refs{\CIV,\CV} 
with the $U(1)$ gauge field that is left over after the $SU(N)$ factor of the gauge group confines.  This suggests that we should simply {\it identify} Fayet-Iliopoulos D-terms on the two sides.
More precisely, the Lagrangian of the theory after the transition can be
written in terms of ${\cal N}=1$ superfields,
\eqn\SS{
\CS = S + \sqrt{2}\theta \psi + \theta \theta F,
}
\eqn\gg{
W_{\alpha} = -i\lambda_{\alpha} + \theta_{\alpha} D + 
{i\over 2} (\sigma^{\mu \nu} \theta)_{\alpha} F_{\mu \nu}
}
as an ${\cal N}=2$ action deformed to $\CN=1$ by the
superpotential \suppp ,
\eqn\action{
{\cal L} = {1\over 4\pi}{\rm Im}\Bigl( \int 
d^2 \theta d^2 {\bar \theta} \;{\bar \CS}{\del {\cal F}_0 \over \del \CS} + 
\int d^2 \theta \; {1\over 2}{\del^2 {\cal F}_0 \over \del \CS^2}\, 
W^{\alpha} W_{\alpha} + 2\int d^2\theta\; {\cal W}(\CS)\Bigr).
}
The Fayet-Iliopoulos deformation \DD\ should produce an 
additional term in this Lagrangian, 
\eqn\da{
\Delta{\cal L} = {j\over 2 \sqrt{2}\pi g_s} D,
}
corresponding to an FI term as in \FI .
Note that on the D-brane side, the center of mass $U(1)$ corresponds to $1/N$ times the identity matrix in $U(N)$, so that the normalization of
\da\ precisely matches \DD .

We now show that the deformation \da\ leads to precisely the 
physics that we expect on the basis of large $N$ duality. 
After turning on the FI term, the effective potential of 
the theory becomes
\eqn\pot{
V = {1\over 4\pi} G^{S{\bar S}}\left(
|\del_{S} {\cal W}|^2 + |j/g_s|^2\right) + \,const.
}
where 
$$
G_{S{\bar S}} = {1\over 2 i } (\tau - {\bar \tau}),\qquad\qquad \tau = {\del^2 \over \del S^2} {\cal F}_0.
$$
We have also shifted the potential by an (arbitrary) constant,
which we choose to be the tension of the branes 
at vanishing $j$,
$$
const. = N {b_{NS} \over 2 \pi g_s},
$$ 
for convenience. We can then rewrite \pot\ as 
\eqn\potwo{{\eqalign{
V & = {i \over 2\pi( \tau - {\bar \tau})} \Bigl(|\alpha - N \tau|^2 + 
|{j/ g_s}|^2\Bigr) + \, const. \cr
&= 
{i \over 2\pi( \tau - {\bar \tau})} 
|{\tilde \alpha} - N \tau|^2+ \,const.
}}
}
where
$$
\tilde{\alpha} = b_{RR} + {i \over g_s} \sqrt{b_{NS}^2+j^2}
$$
and the constant has shifted.  As expected from the D-brane picture,
the effective potential of the theory with the FI 
term turned on and with gauge coupling \first\ is the same as 
that of the theory without the FI term and with gauge coupling \second .  

In this simple example, the prepotential is known to be given exactly by
$$
2\pi i\CF_0(S)={1\over2}S^2\left(\log({S\over\Lambda_0^2 m})-{3\over2}\right).
$$
The vacuum of the theory is determined by the minimum of \potwo, which occurs at
\eqn\vacz{{\tilde\alpha} - N \tau =0,}
or, in terms of the expectation value of the gaugino bilinear, at
\eqn\vac{
S_{*} = m \Lambda_0^2\, \exp(2\pi i {\tilde \alpha} / N).
}
Finally, we note that the energy in the vacuum \vac\ is larger than 
that in the $j=0$ vacuum by the constant that enters \potwo, so 
\eqn\en{
V_{*} = {N}
{\sqrt{b_{NS}^2+j^2}\over 2\pi g_s} 
,
}
which is precisely the tension of the brane
after turning on $j$. This is a strong indication that we
have identified parameters correctly on the two sides of the duality.

It is easy to see that in the vacuum, neither the F-term
$$
\del_S {\cal W} \neq 0,
$$
nor the D-term vanishes. Nevertheless, as will now show, this new
vacuum preserves half of the ${\cal N}=2$ supersymmetry of the theory
we started with, though not the one manifest in the action as written.  
Defining the $SU(2)_R$ doublet of fermions
$$
\Psi = \pmatrix{\psi\cr \lambda},
$$
the relevant part of the supersymmetry transformations of 
the ${\cal N}=2$ theory are 
$$
\delta \Psi^i =  {X}^{ij} \epsilon_j
$$
where ${X}$ is a matrix of F- and D-terms, shifted by an
imaginary part due to the presence of a ``magnetic'' FI term (see
for example \ABSV\ and references therein)
\eqn\XX{
{X} = {i\over\sqrt{2}}
\pmatrix{-Y_1-iY_2+ N & Y_3\cr Y_3 & Y_1-iY_2+N}
}
where the $\CN=2$ auxiliary fields are identified with the auxiliary F-term of 
${\cal S}$ in \SS\ and the D-term of the 
gauge field in \gg\ according to
$$(Y_1+iY_2)=2iF\qquad Y_3=\sqrt{2}D.$$
Note that the
triplet ${\vec Y} = (Y_1,Y_2,Y_3)$ transform like a vector of the
$SU(2)_R$ symmetry of the ${\cal N}=2$ theory.
In the vacuum \vacz\  
$$
{X} =
{iN\over \sqrt{2(b_{NS}^2+j^2)}} 
\pmatrix{b_{NS}-\sqrt{b_{NS}^2+j^2}
 & j\cr j & 
-b_{NS}- \sqrt{b_{NS}^2+j^2}}.
$$
The supersymmetry manifest in \action\ corresponds to $\epsilon_1$,
and it is clearly broken in the vacuum for nonvanishing $j$, 
since neither the F- nor the D-term vanish. However, the 
determinant of ${X}$
vanishes, and so there is a zero eigenvector 
corresponding to a preserved supersymmetry.

So far, we have identified turning on $j$ with turning on an FI term
in the low-energy effective action. It is natural to ask what this corresponds
to geometrically in the Calabi-Yau manifold?
In \refs{\TV,\V} (following \refs{\PS,\Mich})
it was shown that turning on a subset of the 
FI terms of the low-energy ${\cal N}=2$ theory arising 
from IIB compactified on a  Calabi-Yau manifold
corresponds to turning on $H_{NS}$ and $H_{RR}$ fluxes in 
the geometry. This is what we used in \suppp . 
The question of what corresponds to introducing 
the full set of FI terms allowed by ${\cal N}=2$ supersymmetry
was studied, for example, in \refs{\LM,\Hitchin}.
To make the $SU(2)_R$ symmetry of the theory manifest, we can 
write the triplet of the ${\cal N}=2$ FI terms as
$$
E = {i\over\sqrt{2}}\pmatrix{-E_1-iE_2 & E_3\cr E_3 & E_1-iE_2}
$$  
where $(E_1,E_2,E_3)$ transform as a vector under $SU(2)_R$
and enter the action as 
$$
{1\over 4\pi}Re({\rm Tr}{X \bar E})
$$
These are given in terms of ten dimensional
quantities by\foot{This follows from equation $(3.53)$ of \Hitchin\ up to
an $SU(2)_R$ rotation and specializing to a local Calabi-Yau. More
precisely, to derive this statement one needs to look at the the
transformations of the ${\cal N}=2$ gauginos, not the gravitino as in
\Hitchin , but these are closely related. See, for example, \Mich .}
$$
 E_1 = \int_{B} H_{NS}/g_s, \qquad  E_2 = \int_{B} H_{RR}, 
\qquad  E_3 = \int_{B} dJ/g_s.
$$

Note that this agrees precisely with what we have just derived using
large $N$ duality.  Just as the bare gauge coupling
$\int_{S^2} B_{NS}/g_s = b_{NS}/g_s$ gets mapped to
$\int_{B}H_{NS}/g_s$ after the transition due to running of the
coupling, $SU(2)_R$ covariance of the theory demands that turning on
$\int_{S^2} J/g_s = j/g_s$ before the transition get mapped to turning
on $\int_{B}dJ/g_s$ after the transition. Moreover, we saw in this
section that the latter coupling gets identified as a Fayet-Iliopoulos
D-term for the $U(1)$ gauge field on the gravity side. This exactly
matches the result of \refs{\LM,\Hitchin}, since $E_3$ is the
Fayet-Iliopoulos D-term parameter. It is encouraging to note that \refs{\LM,\Hitchin} reach this conclusion via arguments completely orthogonal to ours.
Finally, we observe that an $SO(2)\subset SU(2)_R$ rotation can be used to
set the Fayet-Iliopoulos D-term $E_3=j/g_s$ to zero, at the 
expense of replacing 
$E_1=  b_{NS}/g_s$ by $E_1 =\sqrt{b_{NS}^2+j^2}/g_s$, and this  
directly reproduces \potwo.

\newsec{An $A_2$ Fibration and the Geometric Engineering of 
a Metastable Vacuum}

By wrapping D5 branes on rigid ${\IP}^1$'s in more general geometries with
$b_2(X)>1$, we can engineer vacua which are guaranteed to be massive and break supersymmetry by choosing the central charges of the
branes to be misaligned. 
Since the D-brane theories experience confinement and gaugino
condensation at low energies, we expect to be able to study the dynamics of these vacua
in the dual geometries where the branes are replaced by fluxes.  

In this section, we'll 
consider the simple example of an $A_2$ singularity
fibered over the complex plane $~\IC[t]$.
This is described as a hypersurface in $~\IC^4$,
\eqn\ALE{
uv=z(z-m t)(z-m(t-a)).
}
This geometry has two singular points at $u,v,z=0$ and $t=0,a$.
The singularities are isolated, and blowing them up replaces each with
a rigid $\IP^1$. 
The two ${\IP}^1$'s are independent in
homology, and the local geometry near each of them is the same as that 
studied in the previous section.

Consider now wrapping $N_1$ D5 branes on the ${\IP}^1$ at
$t=0$ and $N_2$ branes on the ${\IP}^1$ at $t=a$.  If the central
charges of the branes,
\eqn\centwo{
Z_i = \int_{S^2_i} J+ i B_{NS} = j_i+ i b_{NS,i}
}
are aligned (e.g., if the K\"ahler parameters $j_i$ both vanish),
the theory on the branes has ${\cal N}=1$ supersymmetry.
At sufficiently low energies, it reduces to a 
$U(N_1)\times U(N_2)$
gauge theory with a bifundamental hypermultiplet $Q,{\tilde Q}$,
a pair of adjoint-valued chiral fields $\Phi_{1,2}$ 
and a superpotential given by
\eqn\treel{
W={m\over 2} Tr \Phi_1^2-  {m\over 2} Tr \Phi_2^2-
a Tr Q{\tilde Q}+
Tr(Q \Phi_1 {\tilde Q} - Q{\tilde Q} \Phi_2).
}
For a small relative phase of the central charges, e.g., by
deforming the theory at vanishing K\"ahler parameters 
$j_i$ by 
\eqn\glim{
j_i/b_{NS,i} \ll 1,
}
we expect this to have a pure gauge theory description at low energies in terms of the supersymmetric theory with Fayet-Iliopoulos terms for the two $U(1)$'s.

Misaligning the central charges such that
\eqn\condi{
Z_1\neq  c_{12} Z_2,
}
for any positive, real constant $c_{12}$, should break all the
supersymmetries of the background. Nevertheless, for large enough $m$
and $a$, the vacuum should be stable. Since the theory is massive, we
expect it to exhibit confinement at very low energies, with broken
supersymmetry. Nevertheless, as we'll now argue, the dynamics of the
theory can be studied effectively for any $j_i$ in the dual geometry,
where the branes have been replaced by fluxes.

\subsec{Large $N$ dual geometry}
 
The Calabi-Yau \ALE\ has a geometric transition which 
replaces the two ${\IP}^1$'s by two $S^3$'s,
$$\quad\qquad\qquad\qquad~~~~~~ S^2_i \qquad \rightarrow \qquad S^3_i \qquad\qquad\qquad i=1,2.$$
The complex structure of the geometry after the transition is encoded in its description as a hypersurface,
\eqn\ALEb{
uv=z(z-m t)(z-m(t-a)) + c t + d,
}
where $c,d$ are related to the periods, $S_{1,2}$, of the 3-cycles,
$A_{1,2}$, corresponding to the two $S^3$'s,
$$
S_{i}  = \int_{A_i} \Omega, \qquad {\del \over \del S_i} {\cal F}_0 = \int_{B_i} \Omega.
$$
As before, $B_i$ are the noncompact 3-cycles dual to $A_i$, and ${\cal F}_0$ is the prepotential of the ${\cal N}=2$ theory. 
The prepotential in this geometry is again given by an exact formula,
$$
2 \pi i {\cal F}_0 =
{1\over 2}\, S_{1}^2\, 
(\log({S_1 \over \Lambda_0^2 m}) - {3\over 2}) 
+
{1\over 2}\, S_{2}^2\, 
(\log({S_2 \over \Lambda_0^2 m}) - {3\over 2}) 
- 
S_1 S_2\, \log({a\over \Lambda_0}).
$$ 
The theory with $N_i$ D5 branes on the ${\IP}^1_{i}$
before the transition is dual to a theory 
with $N_i$ units of RR flux through 
$S^3_i$ after the transition:
$$
\int_{A_i} H_{RR} = N_i.
$$
There are additional fluxes turned on
through the noncompact, dual $B$-cycles,
$$
{\alpha}_i = \int_{B_i} (H_{RR} + 
i{H_{NS}/g_s}) = b_{RR,i} + i {b_{NS,i}/ g_s},
$$
corresponding to running gauge couplings, and
$$
\int_{B_i} {dJ/ g_s}  =  {j_i/ g_s}
$$ 
corresponding to Fayet-Iliopoulos terms. 
The fluxes generate a 
superpotential,
$$
{\cal W} = \int_X \Omega\wedge(H_{RR} + i{H_{NS}/ g_s})
$$
or
$$
{\cal W} = \sum_{i} \,\alpha_iS_i  - N_i {\del \over \del S_{i}} {\cal F}_0, 
$$
and Fayet-Iliopoulos D-terms,
$$
\Delta\CL=\sum_i {j_i\over 2 \sqrt{2} \pi g_s} D_i,
$$
where $D_i$ are auxiliary fields in the two ${\cal N}=1$ vector multiplets. 

Large $N$ duality predicts that for misaligned central charges 
\condi, 
the fluxes should break all supersymmetries, and moreover, that the non-supersymmetric vacuum should be metastable. We'll now show that this indeed the 
case.  The tree-level effective potential of the theory is
$$
V = {1\over 4 \pi } G^{i{\bar k}} \Bigl({\del}_{i}{\cal W} 
{\overline{{\del}_{k}{\cal W}}}+ 
{j_i j_k / g_s^2}\Bigr) +  const,
$$
where 
$$
G_{i{\bar k}} = {1\over 2 i}(\tau - {\bar  \tau})_{ik} \qquad \tau_{ik} = 
{\del^2\over \del S_i\del S_k} {\cal F}_0,
$$
and the K\"ahler metric is determined by the off-shell ${\cal N}=2$ supersymmetry of the background.
We have shifted the zero of the potential energy 
by the tension of the branes at vanishing $j_i$, 
$$
const. = \sum_{i=1,2}N_i {b_{NS,i}\over 2 \pi g_s}.
$$
In the case at hand,
$$
\tau_{11} = {1\over 2 \pi i}\log({S_1\over\Lambda_0^2 m}), \qquad 
\tau_{22} = {1\over 2 \pi i}\log({S_2\over\Lambda_0^2 (-m)}), \qquad 
$$
whereas $\tau_{12}$ is a constant\foot{For convenience, we will take $\tau_{12}$  
to be purely imaginary.} independent of the $S_i$,
$$
\tau_{12} = - {1\over 2 \pi i} \log(a/\Lambda_0).
$$
It is straightforward to see that the critical points of the potential
correspond to solutions of
$$\eqalign{
Re(\alpha_i)+Re(\tau_{ik})N^k=&\,0\cr
G^{ji}G^{jk}\left(Im(\alpha_i)Im(\alpha_k)+j_ij_k/g_s^2
\right)&=(N^j)^2.
}
$$
The first equation fixes the phase of $S_i$'s, and the second their
magnitude. 
Consider the case where the two nodes are widely separated, namely,
where the sizes $S_{i}$ of the two $S^3$'s are much smaller than the
separation $a$ between them.
In this limit, the equations of motion can be easily solved to obtain
\eqn\expec{
\eqalign{
S_{1,*}^{N_1} =& (\Lambda_0^2 m)^{N_1} 
({a\over \Lambda_0})^{N_2 \cos{\theta_{12}}}
\exp(2 \pi i {\tilde \alpha}_1 )+\ldots\cr
S_{2,*}^{N_2} =& (-\Lambda_0^2 m)^{N_2} 
({a\over \Lambda_0})^{N_1 \cos{\theta_{12}}}
\exp(2 \pi i {\tilde \alpha}_2 )+\ldots}
}
where 
$\theta_{ij}$ 
is the relative phase between the central charges
$Z_i$ and $Z_j$.  We can see that in the limit where the $Z_i$ are
aligned, this reduces to the simple case without FI terms where the
effective gauge coupling has been replaced with the parameter
${\tilde \alpha}_i$. The case of anti-aligned central charges was studied
in \refs{\ABSV,\ABF}.
The weak coupling limit of two widely separated nodes, in which our approximation is justified, corresponds to
\eqn\approxi{
S_{i,*} \ll a <\Lambda_0.
}
$S_{i,*}$'s should be identified with the vev's of gaugino condensates
on the branes, and are the order parameters of the theory. This is the
case even in the presence of FI terms, as explained in the previous
section. For small FI terms, this relies only on the off-shell ${\cal
N}=1$ supersymmetry of the theories on both sides of the duality and
a comparison of superpotentials. In \ABSV\ it was conjectured that this
also holds in the brane/antibrane case, where the central charges are anti-aligned and supersymmetry is maximally broken.  It is natural, then, that the
above limit should correspond to the the theory being weakly coupled at
the scale of the superpotential
\treel.  

In the same limit, the vacuum energy is given by
\eqn\vvac{
V_{*} =
N_1 {\sqrt{b_{NS,1}^2+j_1^2}\over 2\pi g_s}  + 
N_2 {\sqrt{b_{NS,2}^2+j_2^2}\over 2 \pi g_s} + 
{1\over 4 \pi^2} N_1N_2 \log({a\over \Lambda_0}) (1-\cos\theta_{12}) +\ldots
}
Note that in the limit of aligned central charges, the
potential energy is simply the brane tension. This is in fact true
$exactly$, and is related to the fact (which will demonstrate later
on) that in this case supersymmetry is preserved.  
For any other value of the angle, there is an additional attraction.
In the extreme case, when we increase $\theta_{12}$ from zero to
$\pi$, we end up with a brane/antibrane system on the flopped
geometry. We can view this as varying one of the $Z_i$'s until the
$B_{NS}$ field through that cycle goes to $minus$ itself. 
\bigskip
\centerline{\epsfxsize 4.0truein\epsfbox{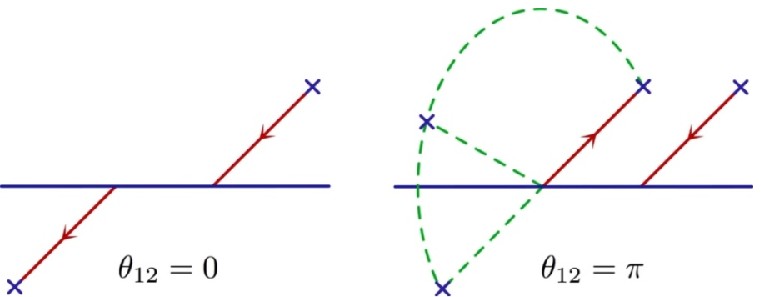}}
\noindent{\ninepoint
\baselineskip=2pt {\bf Fig. 1.} {The $A_2$ geometry from in the text, drawn in the $T$-dual NS5 brane picture. The D5 branes map to D4 branes and appear as red lines. The NS5 branes are drawn as blue lines/points. At $\theta_{12} =0$ the system is supersymmetric. For any other value of $\theta_{12}$, supersymmetry is broken. Varying 
$\theta_{12}$ continuously from zero to $\pi$ produces a geometry which is related to the original $A_2$ geometry by a flop.  }}
\bigskip
\noindent This is a flop, and by comparing to 
\ABF, it follows that the solution we found above for $\theta_{12} = \pi$ 
precisely corresponds to a brane/antibrane system in the 
flopped geometry.

To see that supersymmetry is broken by the vacuum at nonvanishing 
$\theta_{12}$, we
write the action \action\ in an ${\cal N}=2$ invariant way in terms of ${\cal N}=2$
chiral multiplets ${\cal A}_i$
consisting of ${\cal N}=1$ chiral multiplets ${\cal S}_i$,
and ${\cal W}_{i}^{\alpha}$,
$$
{\cal A}_i = ({\cal S}_i, {W}^{\alpha}_i)
$$
or
$$
{\cal A}_i = {S}_i + \theta^a \Psi_{a,i} + \theta^{a}\theta^{b} X_{ab,i} +
{1\over 2} \epsilon_{ab} (\theta^a \sigma^{\mu\nu} \theta^b) F_{\mu\nu}+\ldots.
$$
The appropriate $\CN=2$ Lagrangian is given by
$$
\CL={1\over 4 \pi} {\rm Im}\Bigl(\int d^4\theta d^4x~\CF_0(\CA_i)\Bigr) +{1\over4\pi}Re(X_{i}^{ab} \bar E^i_{ab})
$$
where $X_i^{ab}$ is defined as in \XX .
Then, the relevant supersymmetry variations of the fermions are given by
\eqn\svar{
\delta_{\epsilon} \Psi_i^a = X^{ab}_{i} \epsilon_b,
}
and at the extrema of the effective potential,
\eqn\auxi{ X_i =  {i\over\sqrt{2}}
\pmatrix{N^i +  G^{ik} {\rm Im}(\alpha_{k}) &  -G^{ik} {j_k/g_s}\cr  -G^{ik} {j_k}/g_s &
N^i -G^{ik} {\rm Im}(\alpha_k)}.
}
The equations of motion imply that the determinants of both $X^{1,2}$
vanish. For each node, then, $X^i$ has one zero eigenvalue. It can be
shown (see the general discussion of appendix A), that \svar\ and
\auxi\ imply that a global supersymmetry is preserved if and only if
the central charges
\centwo\ are aligned, i.e., 
if a positive real constant $c_{12}$ exists such that
$$
Z_1 =  c_{12} Z_2.
$$
This is exactly as expected from the open-string picture, and
provides a nice test of the large $N$ duality conjecture for
general central charges.
 
Now we'll show that the vacuum is indeed metastable. Consider the
masses of bosonic fluctuations about the non-supersymmetric vacuum.
As was found to be the case in \ABF, the Hessian of the scalar
potential can be block diagonalized.  After changing variables to
bring the kinetic terms into their canonical form, the eigenvalues
become
\eqn\bosemasses{\eqalign{
M^2_{\phi_{1,2}}&={(a^2+b^2+2abv)\pm\sqrt{(a+b)^2(a-b)^2+4abv(a+b)(b+a)}\over2(1-v)^2}
\cr
M^2_{\phi_{3,4}}&={(a^2+b^2+2abv\cos{\theta})\pm\sqrt{(a+b)^2(a-b)^2+4abv(a+b\cos{\theta})(b+a\cos{\theta})}\over2(1-v)^2}
}}
where we've adopted the notation of \ABSV\ in defining
$$
a={N_1\over2\pi G_{11}|S_{1}|},\qquad\qquad\qquad 
b={N_2\over2\pi G_{22}|S_{2}|},
$$
and
$$
v={G_{12}^2\over G_{11}G_{22}},
$$
with all quantities evaluated in the vacuum.
In addition, 
we've introduced a new angle, which is defined by the equation
$$
\vec{Y}^1\cdot\vec{Y}^2=N_1N_2\cos\theta,
$$
where ${\vec Y}^i$ are related to $X^i$ as in \XX\ for each $i$.
In the limit \approxi, 
$$
\theta=\theta_{12}+\CO(v),
$$
so we can treat this as being the same as the phase which appears in \expec,\vvac.  Note that in the limit
\eqn\susylim{
\cos\theta\rightarrow 1,
}
we recover the results for a supersymmetric system, with the masses of bosons becoming pairwise degenerate. 
The other extreme of anti-aligned charges can be shown to correspond to
$$
\cos\theta\rightarrow -1,
$$
where the results of \ABF\ should be recovered.  
Indeed by plugging in and rearranging
terms we recover the mass formulas from page 25 of \ABF .
These then provide exact values for the tree-level masses of
the component fields in the supersymmetry-breaking vacuum.
For any $\theta$, the 
masses of the bosons are all positive as long as 
$$
v<1.
$$
This, in turn, is ensured as long as the metric on 
moduli space is positive definite in the vacuum.
So indeed, the system is metastable, as expected.

To get a measure of supersymmetry breaking, let's now compare the
masses of the bosons and the fermions in this vacuum.  The fermion
masses arise from the superspace interaction which appears as
$$
{1\over 4\pi}Im\left(\int d^4\theta d^4x~{1\over2}\CF_{ijk}(\Psi^i_a\theta^a)(\Psi^j_b\theta^b)(X^k_{cd}\theta^c\theta^d)\right),
$$
where ${\cal F}_{ijk} = \del_{i}\del_{j} \del_{k} {\cal F}_0$.  For
the geometry in question, the prepotential is exact at one-loop order, and
the third derivatives vanish except when all derivatives are with
respect to the same field.  We can then write the fermion mass
matrices for a given node (and non-canonical kinetic terms) as 
$$
M^{i}_{ab}={1\over 16\pi^2 S_i} \epsilon_{ac}X_{cd}^i\epsilon_{db}
$$ 
Performing a change of
basis to give the fermion kinetic terms a canonical form, we can
diagonalize the resulting mass matrix and obtain the mass eigenvalues.
There are two zero modes, 
$$ M_{\lambda_{1,2}}=0, 
$$
corresponding to two broken supersymmetries. In addition
there are two massive fermions, which we label by $\psi_i$,
\eqn\fermimasses{
M_{\psi_{1,2}}={(a+b)\pm\sqrt{(a-b)^2+2abv(1+\cos{\theta})}\over 2(1-v)}.
}
Note that in the supersymmetric limit \susylim,
the masses of $\psi_1$ and $\psi_2$ match those of
$\phi_{1,3}$ and $\phi_{2,4}$, which have become 
pairwise degenerate. For small misalignment, and large separation of the two nodes,
the mass splittings of bosons and fermions are easily seen to go like
$$
{M^2_\phi - M^2_{\psi} \over M^2_\phi + M^2_{\psi}} \sim v \theta_{12}^2,
$$ 
where $v$ goes to zero in the limit of large separation, and 
$\theta_{12}$ measures the misalignment of the central charges.

\subsec{Gauge theory limit}

In the gauge theory limit \glim, the vacuum energy \vvac\ reduces to
\eqn\vvaca{
V_{*} =\sum_i
N_i {b_{NS,i}\over 2 \pi g_s}(1 + {1\over 2} {j_i^2\over b_{NS,i}^2})
+{1\over 8 \pi^2}N_1 N_2 \log({a\over \Lambda_0}) \Bigl(
{j_1 \over b_{NS,1}} - 
{j_2 \over b_{NS,2}}\Bigr)^2 +\ldots.
}
The first terms are classical contributions, as we saw in section two.
The last term comes from a one-loop diagram in string theory, with
strings stretched between the two stacks of branes running around the
loop.  

To begin with, consider the Abelian case,\foot{Although the
rank of the gauge group is not large in this case, the geometric
transition is still expected to provide a smooth interpolation between
the open- and closed-string geometries. For a recent review, see \ABK
, and references therein.  It is natural to expect that for small
deformations by FI terms that break supersymmetry, the two sides still
provide dual descriptions of the same physics.}
$$U(1)\times U(1),
$$
when the gauge theory has no strong dynamics at low energies.
We should be able to reproduce \vvaca\ directly in the field theory by 
computing the one-loop vacuum amplitude in a theory with FI terms turned on.
We can write the classical F- and D-term potential of the gauge theory as
$$
V_{tree}=|F_{\Phi_1}|^2+|F_{\Phi_2}|^2+|F_{Q}|^2+|F_{\tilde Q}|^2+
{1\over 2}g_{YM,1}^2(|q|^2-|\tilde q|^2-\sqrt{2} \xi_1)^2+
{1\over 2} g_{YM,2}^2
(|\tilde q|^2-|q|^2- \sqrt{2} \xi_2)^2,
$$
where
$$
F_{\Phi_1}=m\phi_1-q\tilde q\qquad F_{\Phi_2}=m\phi_2-q\tilde q\qquad F_{Q}=\tilde q(a+\phi_2-\phi_1)\qquad F_{\tilde Q}=q(a+\phi_2-\phi_1)
$$
and $\phi_{i}$, $q,{\tilde q}$ are the lowest components of the corresponding chiral 
superfields.  The gauge theory quantities are related to those of the string theory construction by
$$
{1\over g_{YM,i}^2} = {b_{NS,i} \over 4 \pi g_s}, \qquad 
\xi_i = {j_i \over 4 \pi g_s}.
$$
The identification between the field
theory FI term and the string theory parameter is expected to hold
only for small $j_i/b_{NS,i}$.  For nonzero $\xi_{1,2}$,
supersymmetry appears to be broken since the two D-term contributions
cannot be simultaneously set to zero with the F-terms. In fact, we
know that if the central charges are aligned, this is just a relic of
writing the theory in the wrong superspace.

For large $m,a$, this potential has a critical point at the origin of
field space. At this point, all the F-terms vanish, and there is pure
D-term supersymmetry breaking. The spectrum of scalar adjoint and
gauge boson masses is still supersymmetric at tree-level, since the
only contribution to the masses in the Lagrangian is the FI-dependent
piece for the bifundamentals. This means that the only relevant
contribution to the one-loop corrected potential is from the
bifundamental fields. The scalar components develop a tree-level mass
which is simply given by\foot{One can easily check that for small $r$, 
the masses agree with what we expect from string theory. The bifundamental matter is the same as for the $0-4$ system, with small $B$-fields turned on
along the D4 branes. See, e.g., \refs{\SW,\BD}.}
\eqn\massq{
m^2_{q}=a^2+ r,\qquad m^2_{\tilde q}=a^2-r
}
while the fermion masses retain their supersymmetric value,
$$
m^2_{\psi_q}=m^2_{\psi_{\tilde q}}=a^2.
$$
We have defined the constant 
\eqn\FII{
r=\sqrt{2} (\xi_2 g_{YM,2}^2 - \xi_1 g_{YM,1}^2) =
\sqrt{2}({j_2\over b_{NS,2}} -{j_1\over b_{NS,1}}).
}
The one-loop correction to the vacuum energy density is given by
$$
V^{(1-loop)}= {1\over 64\pi^2}\left(\sum_b m_b^4\log{m_b^2\over\Lambda_0^2} - 
\sum_f m_f^4\log{m_f^2\over\Lambda_0^2}\right)
,
$$
where $m_{b,f}$ are the boson and the fermion masses, and $\Lambda_0$ is the UV cutoff of the theory. 
The limit in which we expect a good large $N$ dual is when the charged fields 
are very massive, $r \ll a^2$, and at low energies the theory is a pure gauge theory.
Expanding to the leading order in $r/a^2$,
the one-loop potential is then given by
$$
V=V_{tree}+{1\over 16\pi^2}r^2 \log{a\over\Lambda_0}.
$$
We have omitted the $\Lambda_0$ independent terms which correspond to the finite 
renormalization of the couplings in the Lagrangian and are ambiguous.
We see that this exactly agrees with effective potential
\vvac,\vvaca\ as computed in the dual geometry, after the transition. 

In the general, $U(N_1)\times U(N_2)$ case, we have a strongly coupled gauge theory at low energies. Nevertheless, since in the $(N_i,{\bar N}_i)$
sector supersymmetry is preserved,
the one-loop contribution of that sector to the vacuum energy density should vanish beyond the classical contribution. Thus, we expect that only
the bifundamental fields contribute to the vacuum energy at this level.  The one-loop
computation then goes through as in the Abelian case, up to
the $N_1N_2$ factor from multiplicity, once again reproducing the
answer \vvaca\ from large $N$ dual geometry.

\subsec{Relation to the work of \ABK\ }

We close with a comment on the relation to the work of \ABK
, to put the present work in context. The $A_2$ model at hand is the same
as the geometry used to engineer the Fayet model in \ABK. More precisely, the authors there engineered a ``retrofitted'' Fayet model.
The parameter $a$ that sets the mass of the bifundamentals was
generated by stringy or fractional gauge theory instantons, and
thus was much smaller than the scale set by the FI terms, which were taken
to be generic. That resulted in F-term supersymmetry breaking which was
dynamical. 

In the present context, we still have a Fayet-type model,
but we find ourselves in a different regime of parameters of the field theory,
where $r/a^2<1$, with $r$ defined in terms of the FI parameters as in
\FII. Outside of this regime, the vacuum at the origin of field
space, with $Q$ and ${\tilde Q}$ vanishing, becomes tachyonic even in 
the field theory, as can be seen from \massq. Once this becomes the case,
the large $N$ dual presented here is unlikely to be a good description
of the physics. For example, for $N_1=N_2=N$ and $r/a^2>1$, it was found in
\ABK\ that the theory has a non-supersymmetric vacuum where all the
charged bifundamental fields are massive and the gauge symmetry is broken
to $U(N)$. This may still have a description in terms of some dual
geometry with fluxes, but not the one at hand. This may be worth
investigating. 

Thus, unlike the models of \ABK, those considered here break supersymmetry 
spontaneously but not dynamically. It would be nice to find a way to 
retrofit the current models and to generate low scale supersymmetry 
breaking in this context. This would require finding a natural way of 
obtaining small FI terms. The mechanism of \ABK\ does not apply here, since
the terms in question are D-terms and not F-terms. This may be possible in
the context of warped compactifications,\foot{The effects of warping in 
the context of \ABSV\ have been studied in \warped. 
In \granatwo , the authors constructed supergravity solutions 
that were subsequently interpreted in \dymarsky , to correspond to 
D5 branes on the conifold with Fayet-Iliopolous terms turned on, 
i.e. the theory we studied in section two. We thank Y. Nakayama for pointing out to us the latter work.}
and compact Calabi-Yau manifolds, perhaps along the lines of \Yu.

\medskip
\centerline{\bf{Acknowledgments}}
\medskip

We would like to thank T. Grimm, S. Hellerman, S. Kachru,
L. Mazzucato, Y. Nakayama, D. Poland, J. Seo, A. Strominger, and C. Vafa for
valuable discussions. This research was supported in
part by the UC Berkeley Center for Theoretical Physics and NSF grant
PHY-0457317. The research of M.A. is also supported by a DOE OJI
Award, the Alfred P. Sloan Fellowship.

\appendix{A}{Fayet-Iliopoulos Terms for ADE Singularities}

The large $N$ duality we studied in the previous sections should generalize to other ADE fibered geometries. In this appendix we'll demonstrate that the large $N$ dual geometries for these more general spaces have some of the same qualitative features. Consider the ADE type ALE spaces
$${\eqalign{ A_k\;:\qquad& x^2+y^2+z^{k+1}=0\cr
D_r\; :\qquad & x^2 + y^2z
+ z^{r-1} = 0\cr
E_6\;: \qquad & x^2+y^3+z^4 =0\cr 
E_7\;: \qquad &
x^2+y^3+yz^3 =0\cr
E_8\;: \qquad & x^2+y^3+z^5 =0
}} 
$$ 
which are fibered over the complex $t$ plane, allowing the 
coefficients parameterizing the deformations to be $t$ dependent.  The requisite deformations of the singularities are canonical (see \CKV\ and references therein). 
In fibering this over the $t$ plane, the $z_i$ become polynomials $z_i(t)$.
At a generic point in the $t$ plane, the ALE space is smooth, 
with singularities resolved by blowing up $r$ independent 2-cycle classes
$$
S^2_i, \qquad i=1,\ldots r
$$
where $r$ is the rank of the corresponding Lie algebra. 
This corresponds to turning on K\"ahler moduli
$$
Z_i=\int_{S^2_i}(J+iB_{NS})=j_i+ib_{NS,i}.
$$
The 2-cycles $S^2_i$ 
intersect according to the ADE Dynkin diagram of the singularity.
Consider now wrapping $N_i$ D5 branes on the $i$'th 2-cycle class.
The theory on the branes is an $\CN=2$ quiver theory with gauge group
$$\prod_i U(N_i),$$ 
with a bifundamental hypermultiplet 
$Q_{ij}$, ${Q}_{ji}$
 for each pair of nodes connected by a link in the Dynkin diagram. The fibration breaks the
supersymmetry to ${\cal N}=1$ by turning on superpotentials 
$
W_i(\Phi_i)
$
for the adjoint chiral multiplets $\Phi_i$, 
$$
W'_{i}(t) = \int_{S_{i,t}^2} \omega^{2,0},
$$
which compute the holomorphic volumes of the 2-cycles at fixed $t$. The superpotentials
$W_i(t)$ can be thought of as parameterizing the choice of
complex structure of the ALE space at each point in the $t$ plane. 
The full tree-level superpotential of the theory is given by
$$
W = \sum_i Tr W_i(\Phi_i) + \sum_{i<j}  Tr 
(Q_{ij}Q_{ji} \Phi_i-Q_{ij}\Phi_j Q_{ji})
$$
where the latter sum runs over nodes that are linked.
 
For vanishing $j_i$, the structure of the vacua of the theory was computed in \CKV. 
For each positive root $e_{I}$ of the lie algebra, 
$$
e_{I} = 
\sum_{I} n_{I}^i e_i
$$ 
for positive integers $n_I^i$, one gets a rigid ${\IP}^1$ at points in the $t$-plane 
$$
t=a_{I,p},
$$
where
\eqn\pos{
W'_I(a_{I,p}) =\sum_{i} n_{I}^i\,W'_i(a_{I,p})=0.
}
Here $I$ labels the 
positive root and $p$ runs over all the 
solutions to \pos\ for that root.  The choice of vacuum breaks the gauge
group down to
$$
\prod_{I,p} U(M_{I,p})
$$
where 
$$
N_i = \sum_{I} M_{I,p} n_{I}^i.
$$
Turning on generic Fayet-Iliopoulos terms for the $U(1)$ centers
of the gauge group factors,
$$
\Delta\CL=\sum_i{j_i\over 2 \sqrt{2} \pi g_s} Tr D_i,
$$ 
breaks supersymmetry while retaining (meta)stability of the vacuum
as long as $j_i$ is much smaller than the mass of 
all the bifundamentals in the vacuum.

The ALE fibrations have geometric transitions in which each 
${\IP}^1$ is replaced by a minimal $S^3$. 
The leading order 
prepotential ${\cal F}_0$ for all these singularities was 
computed in \VUD , and is given by
\eqn\ADE{\eqalign{
2\pi i {\cal F}_0(S) = &{1\over 2}\sum_{b} \;{S_{b}^2} \,
\Bigl(\log\Bigl({S_{b}\over W''_{I}(a_{b})\,\Lambda_0^2}\Bigr)-{3\over 2}\Bigr)
+
{1\over 2}\sum_{b\neq c}\,{{e_{I(b)}\cdot e_{J(c)}}}\,
S_{b}\,S_{c}\,
\log\Bigl({a_{bc}\over \Lambda_0}\Bigr) + \ldots,
}}
where the sum is over all critical points 
$$
b=(I,p),
$$ 
and $I(b)=I$
denotes the root $I$ to which the critical point labeled by $b$
corresponds.  We are neglecting cubic and higher order terms in the
$S_{I,p}$, which are related to higher loop corrections in the open
string theory.  Above, $e_{I}\cdot e_{J}$ is the inner product of two
positive, though not necessarily simple, roots.  Geometrically, the
inner product is the same as minus the intersection number of the
corresponding 2-cycles classes in the ALE space.
In addition, there are fluxes turned on in the dual
geometry which are determined by holography:
$$\eqalign{
\int_{A_{a}}H_{RR}=&M_a\cr
\int_{B_a}(H_{RR}+ {i\over g_s}H_{NS})=b_{RR,I(a)}+{i\over g_s} &b_{NS,I(a)}~~,
\qquad \int_{B_a}dJ=j_{I(a)}.
}
$$ 
The theory on this geometry without fluxes is an $\CN=2$, 
$U(1)^k$ gauge theory, where $k$ is the number of $S^3$'s.  
The effect of the fluxes on the closed-string theory in
this background was determined in \refs{\LM,\Hitchin}. 
The result is a set of
electric and magnetic $\CN=2$ Fayet-Iliopoulos terms, which enter the
$\CN=2$ superspace Lagrangian, 
$$
\CL={1\over 4\pi}{\rm Im}\left(\int d^4\theta ~\CF_0(\CA^a)\right)+
{1\over 4\pi}Re(\vec{Y}^a\cdot\vec{E}_a).
$$
with
$$
\vec{E}_a=\left(~{{b^{NS}_a\over g_s}}~,~b_{RR,a}~,~{{j_a\over g_s}}~\right),
$$
and where the auxiliary fields ${\vec Y}^a$ are shifted
by the magnetic FI term,
$$
\vec{M}^a=(~0~,~M^a~,~0~).
$$
The auxiliary field Lagrangian then has the form
$$
\CL_{aux}={1\over8\pi}G_{ab}Re(\vec{Y})^a\cdot Re(\vec{Y})^b+{1\over 4\pi}Re(\tau_{ab})Re(\vec{Y})^a\cdot\vec{M}^b+{1\over4\pi}Re(\vec{Y})^a\cdot\vec{E}_a
$$
and integrating out the auxiliary fields sets them equal to their expectation values,
$$
\vec{Y}^a=-G^{ab}\left(\vec{E}_b+Re(\tau_{bc})\vec{M}^c\right)+
i\vec{M}^a
$$
or, to be more precise,
$$
-G_{ab}\vec{Y}^b=
\left(b^{NS}_a/g_s~,~b^{RR}_a+\bar\tau_{ab}M^b~,~j_a/g_s\right).
$$
We can make contact with the more familiar form of this action and
its scalar potential by reducing to $\CN=1$ superspace.  There, the
auxiliary fields $\vec{Y}^a$ get identified with auxiliary fields of
the vector and chiral multiplets corresponding to the $a$'th $S^3$,
and the fluxes give rise to the usual flux superpotential 
$$
\CW=\sum_{a}\alpha_aS_a-M_a\del_{S_a}\CF_0(S).
$$
In addition, there are Fayet-Iliopoulos terms for the $U(1)$'s, and the total scalar potential is given by
$$
V={1\over 4\pi} G^{ab}\left(\del_a\CW\overline{\del_b\CW}+j_a j_b/g_s^2\right)
$$
where 
$$\alpha_a=b_{RR,a}+ib_{NS,a}/g_s.
$$  
There are vacua at the field values which satisfy
$$
\del_a V\sim\CF_{abc}G^{ce}G^{bd}\left((\alpha_e-\bar\tau_{em}M^m)(\bar\alpha_d-\bar\tau_{dn}M^n)+j_e j_d/g_s^2\right)=0.
$$
At one-loop order, 
the prepotential has nonvanishing third derivatives only when 
all derivatives are with respect to the same field.
The vacuum condition can be simplified to this order, 
and upon considering the equation as two real equations for the real and imaginary part, the conditions become
$$\eqalign{
\left(b_{RR,a}-Re(\tau_{ab})\right)M^b=&\,0\cr
G^{ac}G^{ad}\left(b_{NS,c}b_{NS,d}+ j_cj_d\right)&=(M_ag_s)^2 .
}$$
The first of these can be solved easily for the phases of the $S^a$.  
Moreover, we see that it is equivalent to the condition that 
the real part of the auxiliary fields $Y_2^a$ vanish for all $a$ 
in the $\CN=2$ superspace Lagrangian,
$$
G_{ab}Re(Y_2^b)=0.
$$  
In light of that result, the second condition can be written as
\eqn\sussy{
\vec{Y}_a\cdot \vec{Y}_a=0.
}
Since the supersymmetry transformations are 
$$
\delta_{\epsilon}\Psi_a =  X_a\epsilon+\ldots
$$
where
$$
X_a = {i\over\sqrt{2}}\pmatrix{-Y_1^a- i Re(Y_2)^a+M^a & Y_3^a\cr Y_3^a & Y_1^a -i Re(Y_2)^a+M^a},
$$
\sussy\ is precisely the condition that there exists 
some supersymmetry transformation on each node which is locally
preserved by the vacuum.  
Of course, for supersymmetry to be conserved globally, 
these supersymmetry transformations must match for all nodes.  
The condition for this to be the case is
$$
{Y_1^a\over M^a}={Y_1^b\over M^b}\qquad\qquad{Y_3^a\over M^a}={Y_3^b\over M^b},
$$
which, along with the requirement that the metric on moduli space be positive definite in the vacuum, requires that
$$
Z_a = c_{ab} Z_b
$$
for a positive, real constant $c_{ab}$.  
This conforms to our intuition from the open-string 
picture that preserving supersymmetry should require that the 
complex combination of the FI terms and gauge couplings should 
have the same phase on each node.

We can also see that the vacuum we just found is metastable,
as we expect based on large $N$ duality. 
Consider the Hessian of the potential,
$$
4\pi\del_{a} \del_c V = {1\over 8\pi^2S_{a}S_{c} g_s^2}
\Bigl(G^{ia}G^{ac}G^{cj}(
b_{NS}^i b_{NS}^j+j_i j_j) - G^{ac}M^a M^c g_s^2\Bigr),
$$
$$
4\pi\del_{a} \del_{\bar c} V = {1\over 8\pi^2S_{a}\bar S_{c}g_s^2}
\Bigl(G^{ia}G^{ac}G^{cj}(
b_{NS}^i b_{NS}^j+j_i j_j) + G^{ac}M^a M^c g_s^2\Bigr),
$$
and similarly for complex conjugates. 
The eigenvalues of the Hessian are manifestly positive 
in the limit where $G_{ab}$ vanishes for $a\neq b$, which corresponds to widely separated nodes, and where the matrix $\del\del V$
is diagonal. Moreover, the determinant of 
the Hessian is strictly positive for any $G_{ab}$,
so the one-loop Hessian remains positive definite for any $G_{ab}$.

Finally, we can compute the value of the vacuum energy in the limit where the branes are far separated.  The relevant limit in this more general case is
$$
S_{a,*}\ll a_{bc} < \Lambda_0.
$$
The vacuum energy is then given by
$$
V_*=\sum_{b}M^b{\sqrt{b^2_{NS,b}+j^2_{b}}\over2\pi g_s}-{1\over 8\pi^2}\sum_{b\neq c}e_{I(b)}\cdot e_{J(c)}M^bM^c\log{a_{bc}\over\Lambda_0}(1-\cos\theta_{bc})
$$
which reduces to the one-loop value in the gauge-theory limit, as in the $A_2$ case.

\listrefs
\bye